\documentclass[aip,jcp,preprint]{revtex4-1}

\usepackage{color}
\usepackage{amsmath}
\usepackage{amssymb}
\usepackage{graphicx}
\usepackage{bm}

\newcommand{\HN}[1]{{#1}}

\begin{document}

\title{
\HN{Nonequilibrium Generalised Langevin Equation for the calculation of
heat transport properties in model 1D atomic chains coupled to two 3D thermal baths
}
}

\author{H. Ness}\email{herve.ness@kcl.ac.uk}
\affiliation{Department of Physics, Faculty of Natural and Mathematical Sciences,
King's College London, Strand, London WC2R 2LS, UK}

\author{L. Stella}
\affiliation{Atomistic Simulation Centre, School of Mathematics and Physics, Queen's
University Belfast, University Road, Belfast BT7 1NN, Northern Ireland,
UK}

\author{C.D. Lorenz}
\affiliation{Department of Physics, Faculty of Natural and Mathematical Sciences,
King's College London, Strand, London WC2R 2LS, UK}

\author{L. Kantorovich}
\affiliation{Department of Physics, Faculty of Natural and Mathematical Sciences,
King's College London, Strand, London WC2R 2LS, UK}

\begin{abstract}
We use a Generalised Langevin Equation (GLE) scheme to study the thermal transport of 
low dimensional systems.
In this approach, the central classical region is connected to two realistic thermal baths 
kept at two different temperatures [H. Ness {\it et al.}, Phys. Rev. B {\bf 93}, 174303 (2016)]. 
\HN{We consider model Al systems, i.e. one-dimensional atomic chains connected to three-dimensional
baths. The thermal transport properties are studied as a function of the chain length $N$ and the 
temperature difference $\Delta T$ between the baths. 
We calculate the transport properties both in the linear response regime and in the 
non-linear regime.
Two different laws are obtained for the linear conductance versus the length of
the chains. 
For large temperatures ($T \gtrsim 500$ K) and temperature differences ($\Delta T \gtrsim 500$ K), 
the chains, with $N > 18$ atoms,
present a diffusive transport regime with the presence of a temperature gradient across the system. 
For lower temperatures($T \lesssim 500$ K)
and temperature differences ($\Delta T \lesssim 400$ K), 
a regime similar to the ballistic regime is observed. 
Such a ballistic-like regime is also obtained for shorter chains ($N \le 15 $). 
Our detailed analysis suggests 
that the behaviour at higher temperatures and temperature differences is mainly due to 
anharmonic effects within the long chains.
}
\end{abstract}


\maketitle

\section{Introduction}
\label{sec:intro}

Understanding the physical properties of low-dimensional thermal transport 
is an active field of research since such properties are often counter
intuitive and present intriguing features \cite{AtleeJackson:1968,Nakawa:1968,
Lepri:2003,Dahr:2006,Dhar:2008,Segal:2016}. 
\HN{A recent review on the subject can be found in  [\onlinecite{Lepri:2016}].}
For instance, some one-dimensional (1D) models violate the well known
Fourier law of heat transport, while some others do not.

It has been known, since the seminal work of Rieder {\it et al.} [\onlinecite{Rieder:1967}],
that in 1D homogeneous harmonic systems (also referred to as
integrable systems), the thermal conductivity diverges in the 
thermodynamic limit. No temperature gradient is formed in the bulk of
the system, since the dominating energy ``carriers'' are not scattered 
and propagate ballistically.
A large variety of harmonic (integrable) 
classical \cite{Rieder:1967,Rich:1975,Spohn:1977,Casati:1979,Hu:2000,Segal:2003,Segal:2008,Kannan:2012,Saaskilahti:2012,Landi:2013}
and quantum \cite{Spohn:1977,Zurcher:1990,Saito:1996,Saito:2000,Dahr:2003,Segal:2003,Gaul:2007,Asadian:2013}
systems have been studied using analytical and/or numerical approaches.
All these studies show that there is no temperature gradient inside
the system (except for small regions in the vicinity of the contacts
between the central system and the baths).
One usually obtains a constant-temperature profile 
\cite{Rieder:1967,Rich:1975,Spohn:1977,Casati:1979,Hu:2000,Kannan:2012,Landi:2013,
Zurcher:1990,Saito:1996,Saito:2000,Dahr:2003,Gaul:2007,Asadian:2013}
in the central system
around the averaged temperature $T_{\rm av}=(T_L + T_R)/2$.
  
On the other hand, in classical or quantum non-integrable 
systems, a temperature gradient is formed inside the system.
The temperature gradient is uniform, and the heat conductivity 
is finite. The transport is said to be diffusive and these systems 
obey Fourier's law.
In order to obtain a diffusive transport regime, one has to
introduce any form of anharmonic effects in the system 
\cite{Jackson:1968,Bolsterli:1970,Nakazawa:1970,Spohn:1977,Eckmann:1999,Hatano:1999,Hu:2000,
Zhang:2002,Pereira:2004,Mai:2006,Bricmont:2007,
Gaul:2007,Segal:2003,Segal:2008,Hu:2010,Giberti:2011,Pereira:2011,Saaskilahti:2012,Shah:2013,Landi:2013}.
Diffusive transport is also obtained when
extra local stochastic processes \cite{Bolsterli:1970,Rich:1975,Spohn:1977,Zurcher:1990,
Pereira:2004,Bernardin:2005,Kannan:2012,Landi:2013}
or extra collision processes \cite{Lepri:2009,Bernardin:2012} are introduced.
A vibrational mode coupling in classical systems \cite{Wang:2004} is also responsible for 
diffusive transport.

The introduction of configurational defects \cite{Jackson:1968,Rubin:1971,Tsironis:1999}  
or disorder \cite{Rubin:1971,Rich:1975,Kipnis:1982,Zhang:2002,Kannan:2012}
in harmonic systems can also lead to diffusive transport, with the build up of a temperature 
gradient across the system.
Defects and disorder introduce some form of localisation of the vibration 
modes \cite{Nakazawa:1970,Stoneham:1975} 
which do not favour ballistic transport \cite{Nakazawa:1970}.

In the previously mentioned studies, the system is either ballistic or diffusive. 
However it is very important
to understand if a crossover between the two regimes can be obtained 
as it has been observed experimentally
in a wide range of low-dimensional systems (from carbon nanotubes \cite{Kim:2001}, 
graphene nanoribbons \cite{Bae:2013}, nanowires \cite{Hsiao:2013} and polymer 
nanofibres \cite{Shen:2010}).
From the theoretical point of view, 
the ballistic to diffusive crossover has been studied phenomenologically by introducing
and tuning a viscosity-like coefficient (in a simple Langevin-like dynamics) on each
atom in the system \cite{Landi:2013,Roy:2008}, by modifying the interatomic
potential \cite{Segal:2003}, or by introducing a simple description of phonon-phonon 
interaction \cite{Yamamoto:2009}.
A single scheme that can describe both regimes is central
and crucially needed to understand the origin of the crossover. 
Recently, a unified microscopy formalism to study the ballistic-to-diffusive crossover was 
provided in \cite{Bagchi:2014} on the basis of a scaling ansatz for model systems.

In this manuscript, we present an application of our recently developed
Generalised Langevin Equation (GLE) method to the study of the heat transport properties
of low-dimensional systems.
Within this single GLE scheme, we simulate the dissipative dynamics of model Al 
systems. 
Our main objective is to determine if different transport regimes can be established
depending on the chain's length and temperatures involved.

We consider 1D atomic chains connected to two realistic thermal 3D baths
kept at temperatures $T_L$ and $T_R$. 
\HN{
By realistic it is meant that (1) the bath are described at the atomic level, 
with a proper 3D structure, (2) the coupling between the system and the baths 
obtained in our GLE approach is not simply given by some arbitrary constant(s), 
(3) the inter-atomic interaction is given by a N-body type interaction 
designed in materials science and not by a model pairwise potential. Such a
N-body potential gives a realistic description of the covalent bonding between 
atoms in metallic systems which is not well described by pair-wise potentials.  
Therefore, we use the Embedded Atom Method (EAM) \cite{Daw:1984,NISTweb,Mishin:1999}
to describe the interatomic potential between the Al atoms in the 1D chains
and between the chains and the two thermal baths. 
}
The only ``free parameters'' that can be changed are the length $N$ of the 1D chains and the 
temperatures $T_L$ and $T_R$ of the baths.
\HN{
We calculate the transport properties of the 1D atomic chains in both the linear reponse regime 
and at full nonequilibrium conditions.
We study the temperature and length dependence of the linear heat conductance and of the
nonequilibrium heat current. We find different transport regimes, which appear to be
more characteristic of either the diffusive or ballistic regime,
depending on the temperatures and temperature
differences and on the length of the chains. We analyse the results in terms of anharmonic
versus harmonic effects in the effective interatomic potential.
}

\section{Method and systems}
\label{sec:method_and_systems}

To study the thermal transport of 1D atomic chains, we used our recently developed GLE
scheme \cite{Kantorovich:2008,Kantorovich:2008b}. 
The method and its implementation in the molecular dynamics package LAMMPS \cite{Plimpton:1995}
have been well documented in recent papers \cite{Stella:2014,Ness:2015}.
The extension of the original GLE scheme to situations where
the central system is interacting with several independent baths has
been recently given in \cite{Ness:2016}. We call it the GLE-2B scheme.
We stress that this method, at least in principle, enables one to obtain an exact solution of 
the problem of heat transport in the classical case. The method is based on a physically realistic 
picture of infinite leads kept at equilibrium (at corresponding temperatures) and
an arbitrary central region which interacts and exchanges heat with the leads. 
The only approximations made are that the leads are treated as harmonic baths and the interaction 
between the leads and the central system is linear in terms of displacements of the atoms in the
leads; the central system is treated without any approximations and can be arbitrarily anharmonic.

{In a recent paper~[\onlinecite{Ness:2016}], we have discussed the advantages of
the GLE-2B approach over other more conventional thermostatting approaches such as 
Nose-Hoover or simple Langevin thermostats. Here, we use our GLE-2B approach as it does not
rely on the use of adjustable parameter(s) to describe the relaxation processes in the baths.
In Ref.~[\onlinecite{Ness:2016}], we have shown that depending upon the value used for such parameters, 
one can obtain completely different physical results. 
In the present work, we want to study exclusively the influence of the length of the system and 
the baths' temperatures on the thermal transport properties. 
}

We now briefly recall the main ``ingredients'' of the method.
We consider a central system (the 1D atomic chain) with a general Hamiltonian dynamics
for the positions $r_{i\alpha}$ and momenta $p_{i\alpha}$ degrees of freedom (DOF) associated
with atom $i$ (mass $m_i$) and Cartesian coordinate $\alpha=x,y,z$.
The central system is connected to two ($\nu=1,2$) harmonic baths, with DOF $u_{b_\nu}$ and
$\dot u_{b_\nu}$, via coupling 
$\sum_{b_\nu} \mu_{l_\nu }f_{b_\nu}(\mathbf{r}) u_{b_\nu}$.
The coupling is linear with respect to the atomic displacement $u_{b_\nu}$ ($b_\nu \equiv l_\nu\gamma$) of the
atom $l_\nu$ ($\gamma=x,y,z$), in bath $\nu$, around its equilibrium position. $\mu_{l_\nu }$ is the mass
of atom $l_\nu$. The coupling force $\mu_{l_\nu }f_{b_\nu}(\mathbf{r})$ between the central system and the bath $\nu$
is arbitrary with respect to the central system DOF $r_{i\alpha}$.
 
By solving Newton's equations of motion for all the DOFs, and integrating out the baths' DOFs,
one obtains the following ``embedded'' dissipative dynamics for the DOF of the central 
system \cite{Kantorovich:2008,Ness:2016}:
\begin{equation}
\begin{split}
m_{i}\ddot{r}_{i\alpha} = \mathcal{F}_{i\alpha}
- & \int_{-\infty}^{t}{\rm d}t' \sum_{\nu, i'\alpha'} K^{(\nu)}_{i\alpha,i'\alpha'}(t,t';\mathbf{r})
	\dot{r}_{i'\alpha'}(t') \\ 
+ & \sum_\nu \eta^{(\nu)}_{i\alpha}(t;{\bf r}) 
\end{split}
\label{eq:GLE}
\end{equation}

The total force $\mathcal{F}_{i\alpha}$ acting on atom $i$ of the central system 
(in direction $\alpha$) arises not only from the interaction between atoms in
the central system, but also from the interaction between these atoms and the atoms
of the harmonic baths which are kept fixed at their equilibrium positions.
The force $\mathcal{F}_{i\alpha}$ also contains a polaronic-like term which
reflects the fact that, due to the coupling between the central system and the
baths, the harmonic oscillators of the baths are displaced
\cite{Kantorovich:2008,Stella:2014,Ness:2015,Ness:2016}.

Eq.~(\ref{eq:GLE}) also contains a generalised memory kernel 
which depends on both times $t$ and $t'$ separately (i.e. not on their difference) 
and also on the spatial coordinates (DOFs) $\mathbf{r}=\left( r_{i\alpha}\right)$ 
of all atoms of the central system. The memory kernel is given by
\begin{equation}
\begin{split}
K^{(\nu)}_{i\alpha,i'\alpha'}(t,t';\mathbf{r})=
\sum_{b_\nu, b'_\nu} \sqrt{\mu_{l_\nu}\mu_{l'_\nu}}\ g_{i\alpha,b_\nu}(\mathbf{r}(t)) \\
\times \Pi_{b_\nu,b'_\nu}(t-t')\ g_{i'\alpha',b'_\nu}(\mathbf{r}(t')) ,
\end{split}
\label{eq:kernel}
\end{equation}
where $g_{i\alpha,b_\nu}$ is the derivative of the coupling force
$g_{i\alpha,b_\nu}=\partial f_{b_\nu}({\bf r})/\partial r_{i\alpha}$
and the polarization matrix $\Pi_{b_\nu,b'_\nu}(t-t')$ is related to the harmonic 
dynamical matrix of the bath $\nu$ [\onlinecite{Kantorovich:2008,Stella:2014,Ness:2015,Ness:2016}].
\HN{In the energy representation, the polarization matrix $\Pi_{b_\nu,b'_\nu}(\omega)$
represents the vibrational density of states of the corresponding bath 
} [\onlinecite{Kantorovich:2008,Stella:2014,Ness:2015,Ness:2016}].
Finally, the terms $\eta^{(\nu)}_{i\alpha}$ contain all the information about the initial 
conditions of the bath $\nu$ DOFs.

By considering these initial conditions as random processes, the corresponding stochastic 
forces $\eta^{(\nu)}_{i\alpha}$ can be described by a multi-dimensional Gaussian stochastic 
process with correlation functions \cite{Kantorovich:2008,Stella:2014}
\begin{equation}
\begin{split}
&\langle\ \eta^{(\nu)}_{i\alpha}(t;{\bf r})\ \rangle = 0 \\
&\langle\ \eta^{(\nu)}_{i\alpha}(t;{\bf r})\ \eta^{(\nu')}_{i'\alpha'}(t';\mathbf{r}) \ \rangle  = \delta_{\nu\nu'}\ k_{B}T_\nu\ 
K^{(\nu)}_{i\alpha,i'\alpha'}(t,t';\mathbf{r}) \nonumber
\end{split}
\end{equation}
and consequently Eq.(\ref{eq:GLE}) corresponds now to a generalised Langevin equation for the central 
system DOFs, with a non-Markovian memory kernel and coloured noise.

\begin{figure}
\includegraphics[width=40mm]{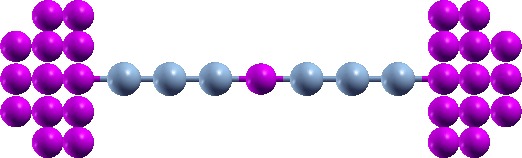}\hspace{15mm}\includegraphics[width=25mm]{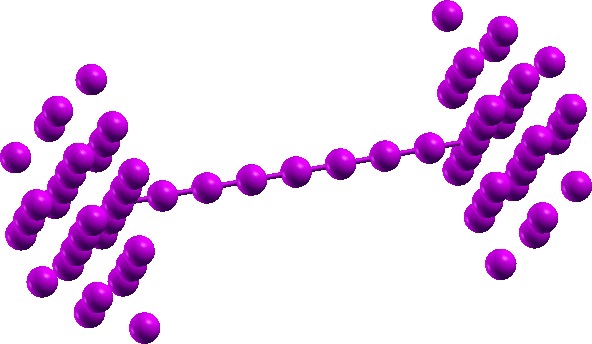}

\vspace{10mm}\includegraphics[width=50mm]{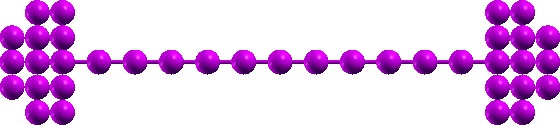}

\vspace{10mm}\includegraphics[width=50mm]{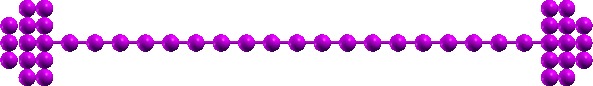}

\vspace{10mm}\includegraphics[width=50mm]{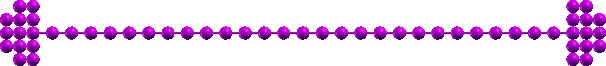}

\caption{Schematic representation of the different one-dimensional Al chains (length $N=7,11,19,27$
from top to bottom) 
connected to two three-dimensional $L$ and $R$ baths (each made of 30 Al atoms). The baths shown
here are in fact what is called the reduced bath regions \cite{Ness:2015,Ness:2016}. These
regions contain atoms $b_\nu$ which interact with atoms $i\alpha$ of the central region via 
non-zero matrix elements $g_{i\alpha,b_\nu}$.
The EAM is used for the interatomic interaction.
The three first (last) atoms in the chains (shown in a different color in the top left corner) are the atoms 
interacting directly with the left (right) bath respectively.}
\label{fig:1Dchains}
\end{figure}

Eq.~(\ref{eq:GLE}) can be efficiently solved numerically by introducing a set of 
auxiliary DOFs (aDOF)
$\{s_{\nu,1}^{(k_\nu)}(t), s_{\nu,2}^{(k_\nu)}(t)\}$  
(the superscript $k_{\nu}$ is used to count the aDOFs)
\cite{Stella:2014,Kupferman:2004,Bao:2004,Luczka:2005,Ceriotti:2009,Ferrario:1979,Marchesoni:1983}
and by mapping the polarization
matrix onto a specific analytical form \cite{Stella:2014} 
\begin{equation}
\begin{split}
\Pi_{b_\nu,b'_\nu}(t-t') \rightarrow 
\sum_{k_\nu} A_{b_\nu b'_\nu}^{(k_\nu)}
e^{-\vert t-t'\vert/\tau_{k_\nu}} \cos(\omega_{k_\nu}\vert t-t'\vert) .\nonumber
\end{split}
\end{equation}
Each pair of aDOF $\{s_{\nu,1}^{(k_\nu)}, s_{\nu,2}^{(k_\nu)}\}$   is associated with
the corresponding mapping coefficients $\{\tau_{k_\nu},\omega_{k_\nu},c_{b_\nu}^{(k_\nu)}\}$.
Then it can be shown that solving Eq.~(\ref{eq:GLE}) is equivalent to 
solving an extended set of Langevin equations, for the central system DOFs and the aDOFs, 
as a multivariate Markovian process, where all the DOF
are independent Wiener stochastic processes with (white noise) correlation functions
\cite{Stella:2014,Ness:2015,Ness:2016}.

The systems we consider are 1D atomic chains (of length $N$) connected to 3D thermal baths
as shown in Figure~\ref{fig:1Dchains}. The electronic transport properties of similar Al 
nanowires have been studied some decades ago \cite{Wan:1997,Taraschi:1998,Taylor:2001}.
It is now interesting to study their thermal transport properties using our method.
We have taken the Embedded Atom Method \cite{Daw:1984} to model the metallic Al system. 
The tabulated interatomic potential, provided by the NIST Interatomic 
Potential Repository Project \cite{NISTweb}, is a typical non-pairwise potential.

{
It is important to note that, when using realistic interatomic interaction potential, 
the model goes beyond the ball-and-spring (with nearest-neighbour interaction) toy models.
In Fig.~\ref{fig:1Dchains} we show that more than two end-atoms of the 1D chains are 
coupled directly to the thermal baths. In the present case, the first (last) 3 atoms of
the 1D chains are coupled directly to the left (right) thermal bath.
Furthermore the coupling of these atoms to the baths is given by the matrix elements
$g_{i\alpha,b_\nu}$ which are not constants throughout the GLE dynamics. These matrix
elements are explicitly dependent on the positions of the atoms in the 1D chains and
change accordingly along the GLE dynamics.

Finally, the thermal baths have their own specific spectral signatures, given by
the polarisation matrix $\Pi_{b_\nu,b'_\nu}(\omega)$ (obtained from the Fourier 
transform
of $\Pi_{b_\nu,b'_\nu}(t-t')$).
The frequency dependence of $\Pi_{b_\nu,b'_\nu}(\omega)$ is not trivial and
depends on the very nature (atomic configuration, chemical nature, etc ...)
of the corresponding bath described at the atomic level. 
Since in our approach we are dealing with a realistic description of the baths,
it is not straight forward to determine if our baths are
ohmic (sub- or super-ohmic) 
as is usually done in other simulations of less realistic models. 
}

\section{Results}
\label{sec:results}

We focus our study on the steady state only and consider chains of 
length $N=7,11,15,19,23,27$ atoms.
To obtain interesting low-dimensional transport properties,
the dynamics of the atoms in the chains is constrained to the 1D motion along the chain axis
($x$-direction) by imposing the condition
$p_{iy}=p_{iz}=0$ for every atom $i$ in the chain.

We calculate the heat current $J_{\rm heat}(N,T_L,T_R)$ from the time derivative of the lead Hamiltonian, 
which results 
in the sum, over the chain atoms, of the products of their velocity and the 
corresponding force \cite{Lepri:2003,Segal:2003} (see Appendix~\ref{app:Jheat}
for further details).

The GLE-2B simulations are typically performed
for 300 ps (150000 timesteps with $\Delta t$ = 2 fs).
The steady state is reached after around 100 ps. We calculate a time average
of the heat current $J_{\rm heat}(N,T_L,T_R)$ over the time range between 200 to 300 ps 
where the system has reached
the steady state (Appendix~\ref{app:Jheat}). 
Furthermore, for information, all the results obtained for the heat current $J_{\rm heat}$
flowing across the chains and for all set of
temperatures $T_L$ and $T_R$
are given in Figure~\ref{fig:Jheat} of Appendix~\ref{app:Jheat_results}.

\begin{figure}
\begin{centering}
\includegraphics[width=85mm]{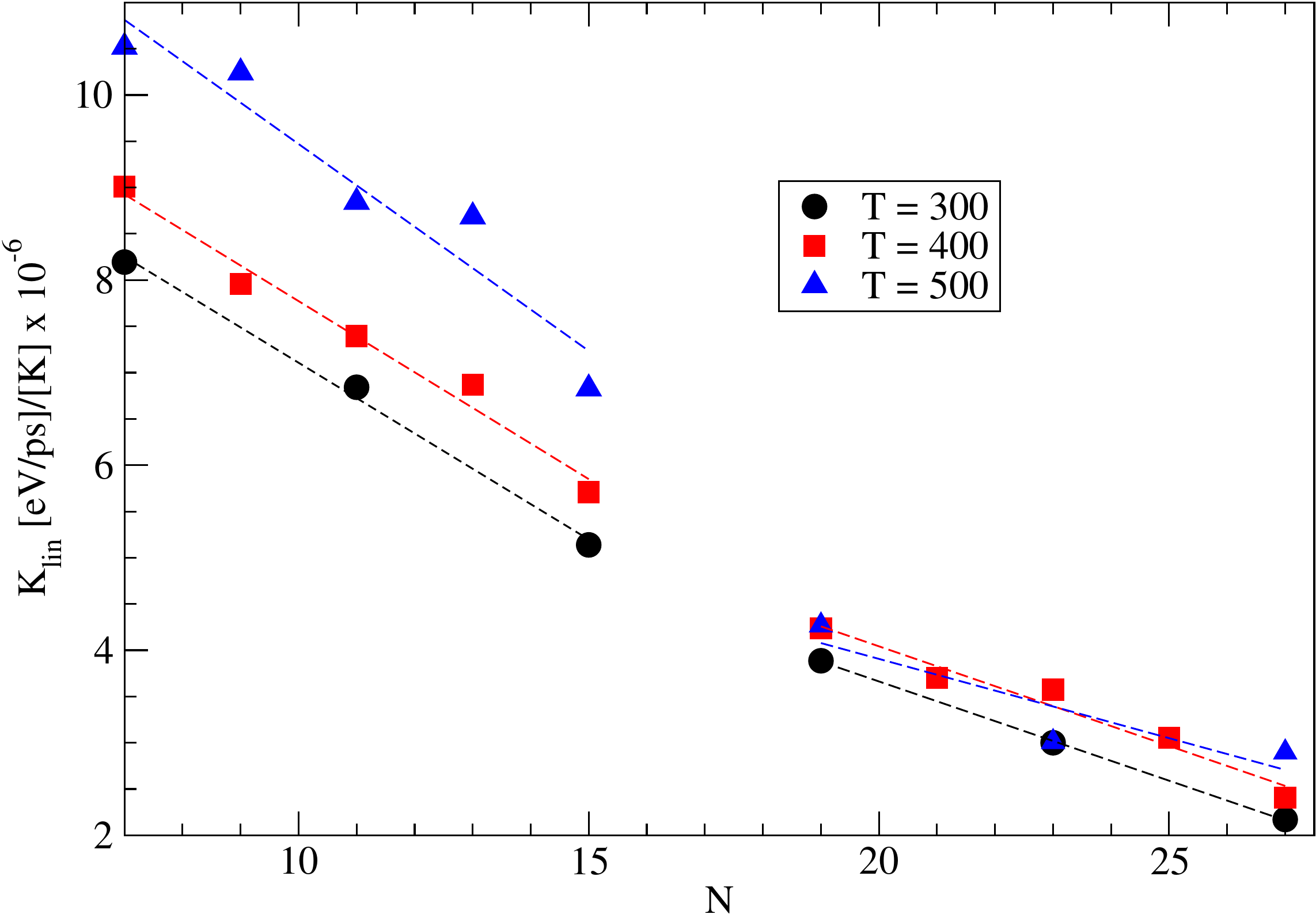}
\end{centering}
\caption{Linear thermal conductance $K_{\rm lin}$ 
for the 1D chains of different length $N$ and for different temperatures $T$.
(Note that, in the linear regime, $\Delta T \rightarrow 0$ and $T_L\rightarrow T_R \equiv T$).
It appears that the length dependence of $K_{\rm lin}$ is not the same for the shorter chains
(i.e. $N < 16$) and for the longest chains ($N > 18$).
\HN{The lines are guides for the eye based on the best linear fit of the calculated points.}
}
\label{fig:Klin}
\end{figure}

From the slope of $J_{\rm heat}(N,T,\Delta T)$ at small temperature difference \cite{Segal:2003}, 
we extract the value of the linear thermal conductance 
$K_{\rm lin}(N,T) = \lim_{\Delta T \rightarrow 0} {J_{\rm heat}(N,T,\Delta T) }/{\Delta T}$.
Figure~\ref{fig:Klin} shows the linear thermal conductance $K_{\rm lin}$
versus the length $N$ for
different temperatures $T_L\rightarrow T_R$.
We can observe two different regimes for the behaviour of $J_{\rm heat}$ versus $N$:
one for short chains (typically $N<16$), and the other for longer chains (typically $N > 18$).

\begin{figure}
\begin{centering}
\includegraphics[width=85mm]{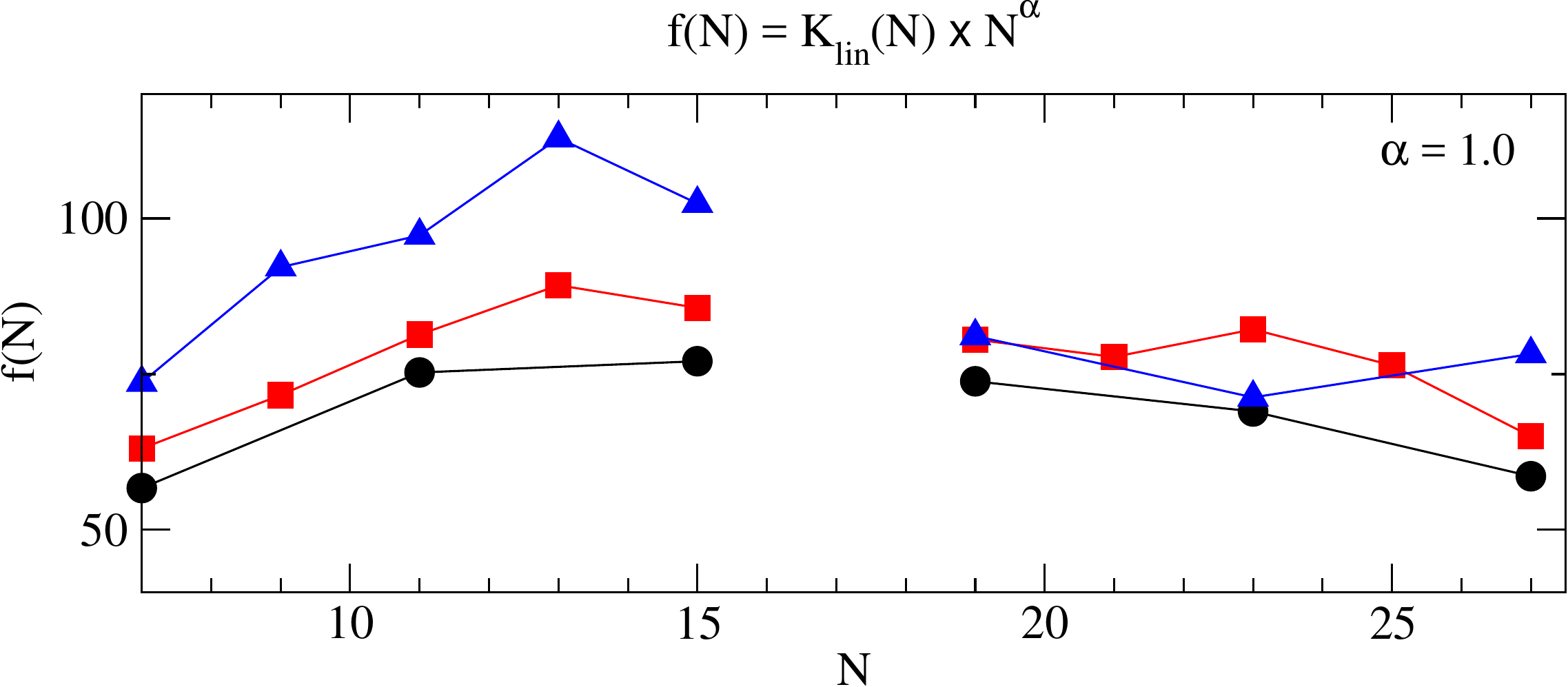}
\end{centering}
\caption{
The function $f(N)=K_{\rm lin}(N,T) N^\alpha$ is plotted for $\alpha=$1.0
for the different temperatures. 
\HN{
It appears that the function $f(N)$ is more constant for the 
longer chains than for the shorter chains.
Hence, for the longer chains, we have $K_{\rm lin}(N) \propto 1/N$ while for
the shorter chains $K_{\rm lin}(N)$ does not follow the same length dependence.
Note that the color of the lines correspond to the temperatures given in Fig.~\ref{fig:Klin}.
The units of $f(N)$ are the same as the unit of $K_{\rm lin}$ in Fig.~\ref{fig:Klin}, 
i.e. $\times 10^{-6}$ [eV/ps K].}
}
\label{fig:Klin_powerlaw}
\end{figure}

For the longer chains, the length dependence of $K_{\rm lin}$ follows quite well the 
typical $1/N$ power law as shown in Fig.~\ref{fig:Klin_powerlaw}.
\HN{We have calculated the the function $f(N)=K_{\rm lin}(N) N^\alpha$ 
for different values of $\alpha \sim 1$, the results for $\alpha=1$ are shown
in Fig.~\ref{fig:Klin_powerlaw} for the different temperatures corresponding
to Fig.~\ref{fig:Klin}.
Qualitatively speaking, the function $f(N)$ is more flat for the longer chains ($N > 18$), 
i.e. $K_{\rm lin}$ follows a power law in $1/N^\alpha$ with $\alpha\sim 1$ (diffusive Fourier law).
For the shorter chains ($N\le15$), the function $f(N)$ clearly presents a stronger dependence
on $N$ (i.e. is not as flat as for the longer chains) implying that $K_{\rm lin}$ follows 
a length dependence law different from $1/N$.
}

\begin{figure}
\begin{centering}
\includegraphics[width=85mm]{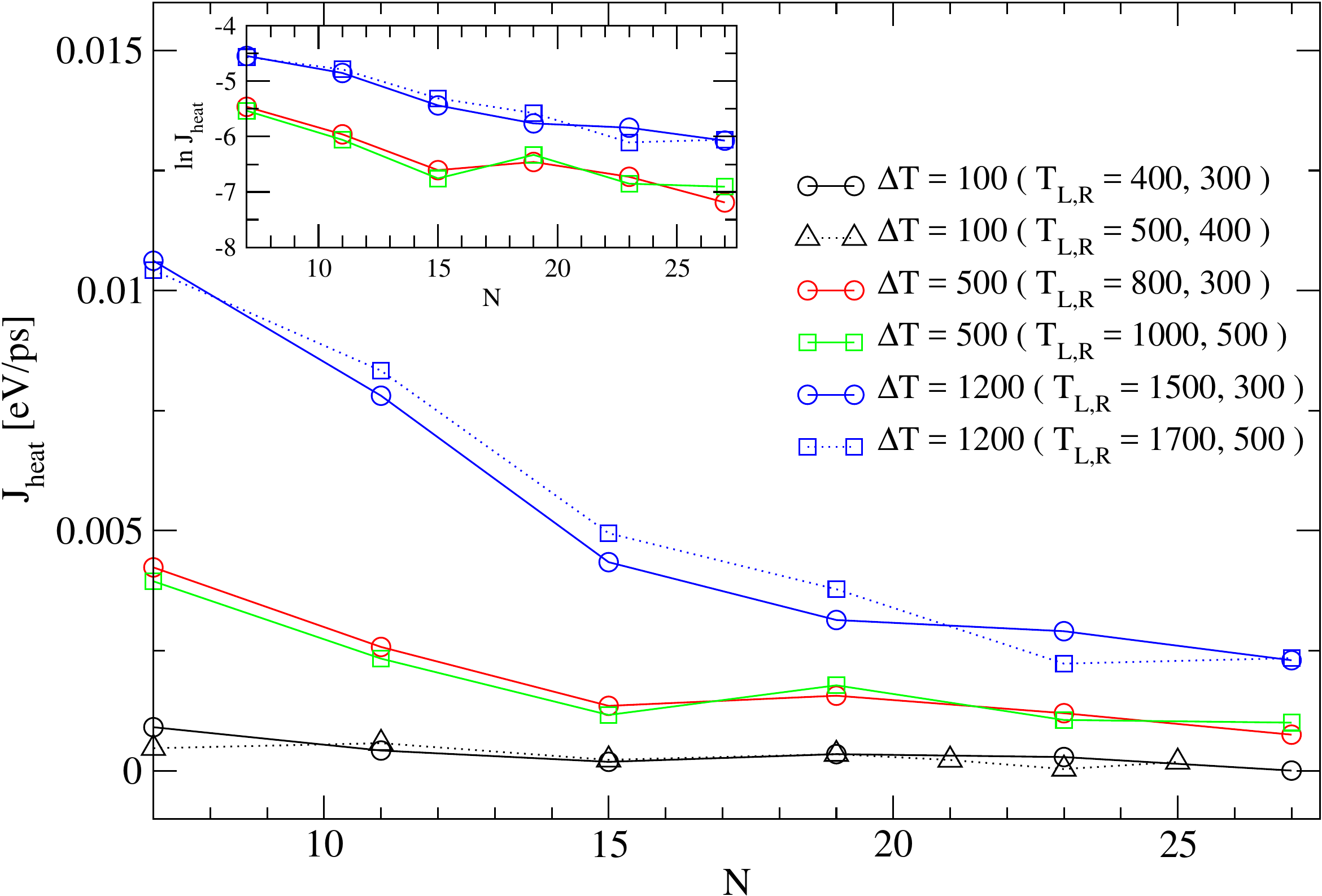}
\end{centering}
\caption{Heat current $J_{\rm heat}$ versus length $N$ of the chains for different sets of bath temperatures
$T_L,T_R$.  For small temperature differences $\Delta T \ll T_L, T_R$, the heat current is almost independent
of the chain length (i.e. ballistic-like regime). 
For larger temperature differences $\Delta T > T_L$ or $T_R$, $J_{\rm heat}$ decreases with the chain
lengths. This behaviour is more typical for the diffusive transport regime. 
{The inset shows that $J_{\rm heat}$, on a logarithmic scale, decreases continuously with increasing $N$
and does not saturate for longer chains.}
}
\label{fig:Jlin_vsL}
\end{figure}

The dependence of the heat current $J_{\rm heat}$ versus length $N$ for different sets of baths' 
temperatures $T_L,T_R$ is shown in Figure~\ref{fig:Jlin_vsL}.
For large temperature differences $\Delta T > T_L$ or $T_R$, the current $J_{\rm heat}$ decreases 
with the chain length which is a characteristic of a diffusive system.
\HN{For such large temperatures and temperature differences, one should keep in mind that
the system is well beyond the linear response regime, and full non-linear and nonequilibrium
effects are present. Determining the true nature of such non-linear effects requires further
investigations which are beyond the scope of the present paper.
}
For small temperature differences $\Delta T \ll T_L, T_R$, the heat current 
appears almost independent of the chain length (such a behaviour would characterise a ballistic transport 
regime). The inset in Figure~\ref{fig:Jlin_vsL} also clearly shows, on a logarithmic scale, that $J_{\rm heat}$ 
does not saturate for the longer chains, especially for larger temperatures and temperature differences 
(the blue curves in Fig.~\ref{fig:Jlin_vsL}).

\begin{figure}
\begin{centering}
\includegraphics[width=85mm]{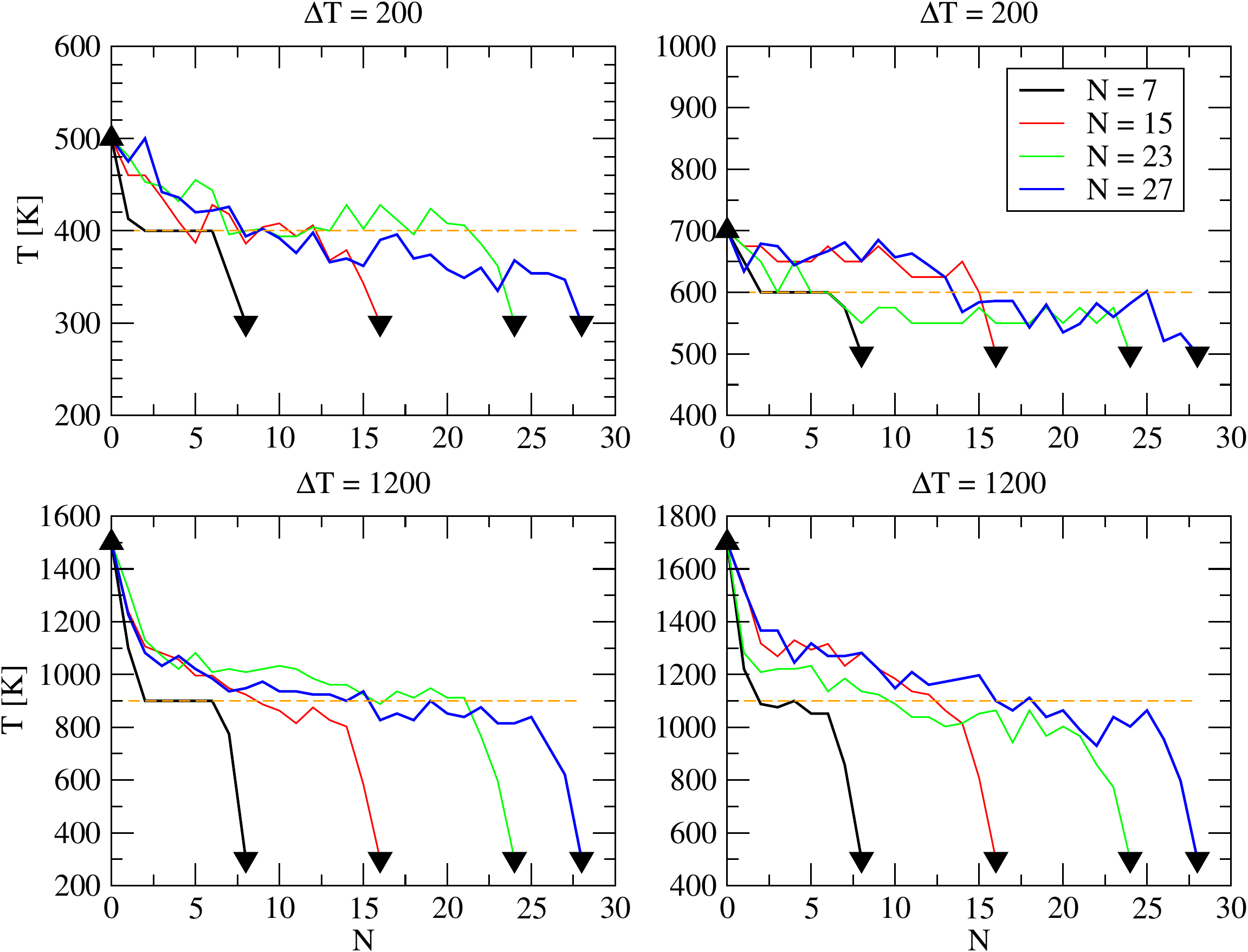}
\end{centering}
\caption{Temperature profiles of 1D chains of different length $N=$ 7,15,23 and 27
for different bath temperatures $T_L$ (filled triangle up) and $T_R$ (filled triangle down).
For the shortest chain ($N=7$), the temperature profile is always flat (except for
the large temperature difference when $T_L=1700$ and $T_R=500$) and is characteristic of the ballistic
transport. For the longest chain $N=27$, the temperature profile in the chain always 
presents a gradient even for $T_L=500$ and $T_R=300$, characteristic of the diffusive transport
regime (Fourier law). An intermediate behaviour is obtained for the other chains $N=15,23$.
}
\label{fig:Tprofile}
\end{figure}

Before we can establish the nature of the transport regime, we now concentrate on the temperature
profiles across the chains.
As we mentioned in the Introduction, the diffusive transport regime
is also associated with the establishment of a linear temperature profile across the
system. The temperature profiles for different chain lengths and different temperature
differences are shown in Figure~\ref{fig:Tprofile}.
One can see the presence of a temperature gradient across the longer chains. The gradient
becomes more pronounced for larger temperature differences. 
[Note that, for the systems considered here (one-dimensional chains connected to 3D baths),
most of the temperature drop occurs at the contact between the chain and the bath (i.e.
large thermal contact resistance) in contrast to systems with a larger (constant and
finite) cross section].

For shorter chains, apart from the case of a very large temperature difference
($T_L=1700, T_R=500$), the temperature profile across the chain is always flat with
the temperature given by  $T_{\rm av}=(T_L + T_R)/2$, see Fig.~\ref{fig:Tprofile}. 
As we mentioned in the Introduction, this behaviour is a typical characteristic
of the ballistic thermal transport regime.
Therefore we can conclude that, by changing the temperature differences and/or the
length of the system, we can obtain two different transport regimes.

The short chains appear to behave like a harmonic system with ballistic transport
properties and no temperature gradient. 
For the longer chains and at high temperature differences, the dependence of 
$K_{\rm lin}(N)$ and $J_{\rm heat}(N)$ vs $N$, and the presence of a
temperature gradient across the chain, indicate a more diffusive transport regime.
This could be a signature either of an anharmonicity in the systems or of the existence 
of localised  vibration modes.
Note that for small temperatures and small temperature differences, all chains appear
to behave as ballistic harmonic systems.

\begin{figure}
\begin{centering}
\includegraphics[width=75mm]{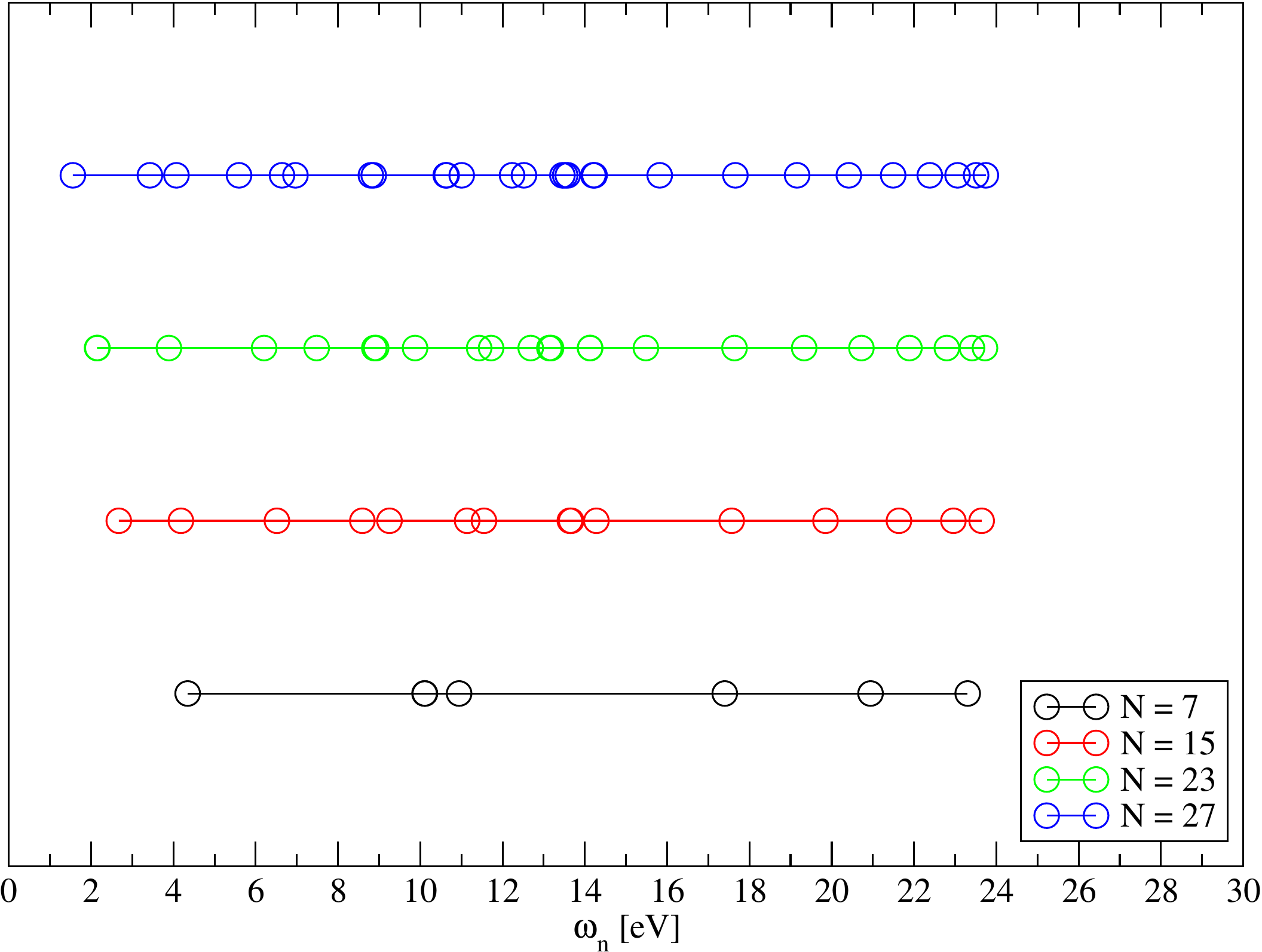}
\end{centering}
\caption{Eigenvalues (in eV) of the vibrational modes
in the chains with $N=7,15,23,27$ (bottom to top lines with circle). More (nearly) degenerate modes
exist in the longer chains. However the ``accumulation'' of a lot of long wave-length modes 
(around $\omega \rightarrow 0$) is not yet achieved for $N=27$. There are no obvious bound states
outside the energy spectrum ($\omega\lesssim 1 $ or $\omega\gtrsim 24$ eV) 
that would correspond to more localised vibrational modes.}
\label{fig:Eigenval}
\end{figure}

It is important now to understand what kind of physics is behind the two transport regimes.
For that we check first how the eigenmodes of vibration of the 1D chains change when $N$
increases. 
The eigenmodes of vibration of the chains are obtained from the diagonalisation of the
dynamical matrix of the chains connected to the baths, as shown on Figure~\ref{fig:1Dchains}.
Figure~\ref{fig:Eigenval} shows the eigenvalues for four different chain lengths. 
One observes more (nearly) degenerate modes in the longer chains and a larger number
of long wavelength modes in the longer chains, as expected. 
Furthermore, we do not see any eigenvalues outside the energy spectrum (i.e.
$\omega\lesssim 1 $ or $\omega\gtrsim 24$ eV) that could correspond to localised modes 
(i.e. bound states).
The presence of such more localised modes (in the longer chains) would lead to the 
breakdown of the ballistic properties.

Another possible mechanism is related to anharmonic effects in the interatomic potential. 
In order to understand such effects, we first consider the work of Segal {\it et al.} \cite{Segal:2003}.
In that work, a simpler bath model and coupling to the bath were used in comparison with 
our GLE-2B. However, in their model calculations, the authors were able to modify the 
interatomic potential used for their description of the interaction in the 1D chains. 
It was found that for purely harmonic chains, 
the heat current is roughly independent of the chain length (a harmonic chain is an integrable model
and presents ballistic properties no matter what the chain length is). 
When anharmonic effects are introduced in the interatomic potential,
$J_{\rm heat}$ becomes length dependent and decreases when chain length increases (see Fig.~10 
in Ref.~[\onlinecite{Segal:2003}]).
Such results are fully compatible with our calculations of $J_{\rm heat}(N)$ shown in
Figure~\ref{fig:Jlin_vsL}.

\begin{figure}
\begin{centering}
\includegraphics[width=85mm]{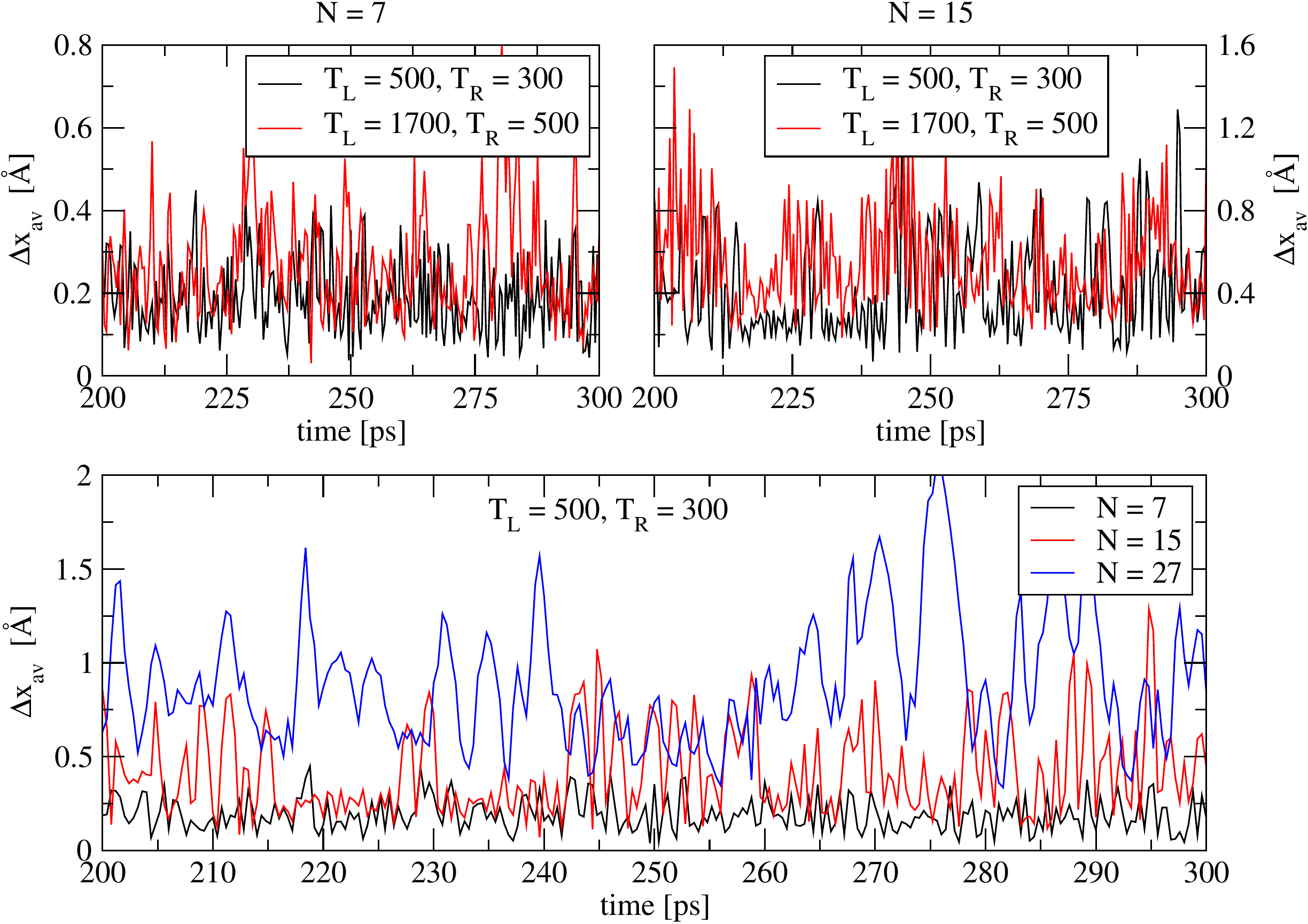}
\end{centering}
\caption{Averaged displacement $\Delta x_{\rm av}$ versus time. 
The value of $\Delta x_{\rm av}$ increases when the temperature
increases (see the two top panels). However such an increase is bigger when one increases the length
of the chains. Hence in short chains, the atoms are more confined around the bottom of the potential 
energy wells, while for longer chains, the atoms have more ``space'' to move and are able to sample 
the anharmonic part of the potential well.}
\label{fig:sqrdispl}
\end{figure}

In order to quantify the presence of the anharmonic effects, we consider the averaged atomic displacement 
$\Delta x_{\rm av}$.
The averaged displacement $\Delta x_{\rm av}(t)$ is obtained
from $\Delta x_{\rm av}(t) = [ \sum_{i=1}^N (\Delta x_i(t))^2 / N ]^{1/2}$ where
$\Delta x_i(t)$ is the relative displacement of atom $i$ in the chain
of length $N$.
The latter is calculated from $\Delta x_i(t) = x_i(t) - \langle x_i \rangle$
where $\langle x_i \rangle$ is the time averaged position of atom $i$
in the time window $[t_{\rm start},t_{\rm stop}]$, i.e.
$\langle x_i \rangle = \sum_{m=0}^M x_i(t_{\rm start}+m \Delta t) / (M+1)$
with $t_{\rm stop}=t_{\rm start}+M\Delta t$.

Figure~\ref{fig:sqrdispl} shows the evolution of $\Delta x_{\rm av}$ versus time when the system
has reached the steady state.
The results show that, for a given chain length, the average variance $\Delta x_{\rm av}$ increases 
when the temperature increases, as expected (top panels in Fig.~\ref{fig:sqrdispl}). 
However, for a given couple of temperatures $T_{L,R}$, the average variance is more pronounced in 
longer chains than in the shorter ones (see the bottom panel in Fig.~\ref{fig:sqrdispl}). 
This suggests that for short chains, the atoms are more confined around the bottom of the 
potential energy wells, while for longer chains, the atoms have more ``room'' to move and 
are able to sample the anharmonic part of the potential energy well. 

\begin{figure}
\begin{centering}
\includegraphics[width=85mm]{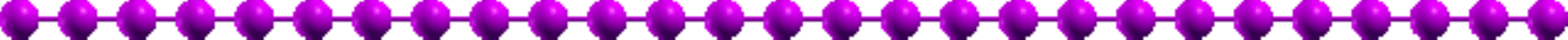}\\
\vspace{5mm}
\includegraphics[width=85mm]{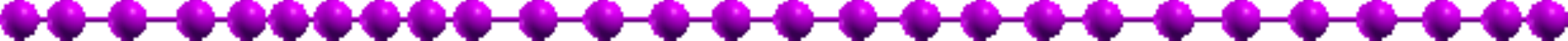}\\
\vspace{5mm}
\includegraphics[width=85mm]{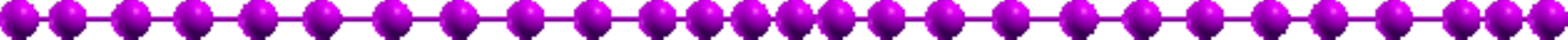}
\end{centering}
\caption{Typical atomic configurations of the long chain with $N=27$ during a GLE run
with $T_L=800$ and $T_R=300$.
The configuration in the top panel corresponds to the initial conditions, with equally spaced
atoms. During the GLE run, a form of clustering appears where the distances between atoms
in (between) the clusters is smaller (larger) than the average interatomic distance,
hence sampling the anharmonic parts of the potential.}
\label{fig:1config_27at}
\end{figure}

Fig.~\ref{fig:1config_27at} shows typical configurations of the atoms in the long chain
with $N=27$. Clearly, during the GLE run, some atoms get closer to each other (with distances
smaller than the average interatomic distance), forming some of kind of clusters. The distance 
between ``clusters'' is also larger than the average interatomic distance.
Such a feature is also characteristic of the presence of long wavelength modes, which are more
numerous in longer chains than in the shorter ones.  

Therefore, we can conclude that, for longer chains and higher temperatures,
the atomic motions sample a larger range of distances and a considerable part of
the anharmonic EAM potential.
This leads to the expected diffusive transport regime
and to the presence of a temperature gradient across the long chains.
However this behaviour is less apparent for low temperatures (as shown for example
for $T_L=400$ and $T_R = 300$ in Fig.~\ref{fig:Jlin_vsL}).

For shorter chains, 
the motion of the atoms is more ``restrained'' and most probably their motion samples 
only the harmonic part of the potential.
This leads to a more harmonic-like system with a more ballistic-like transport regime
with no temperature gradient across the short chains
(except in the regime of very large temperature difference $\Delta T$). 
Note that such a behaviour for the atomic motion could also be ``accentuated'' by the
fact that the three first (last) atoms of the chains are also directly interacting with
the atomic baths.

\section{Conclusion}
\label{sec:ccl}

We have used our recently developed Generalised Langevin Equation with 2 baths (GLE-2B) method \cite{Ness:2016}
to study the thermal transport properties of 1D atomic chains coupled to two realistic 3D thermal baths kept at 
their own temperature. 
The results presented in this paper should be understood as a proof of principle of the
robustness and efficiency of the numerical
GLE-2B methodology that we have recently developed. 

We have found that two different laws are obtained for the linear conductance versus the length of
the chains. Furthermore, for large temperatures and temperature differences, the chains present a diffusive 
transport regime with the presence of a temperature gradient across the system. 
\HN{In such a regime, nonequilibrium effects are present and require an in-depth analysis}.
For lower temperatures
and temperature differences, a regime reminiscent to the ballistic regime is observed. 
In short chains, except for the largest temperature differences considered, the temperature profile 
does not present any gradient, a characteristic of a ballistic transport property.
Our detailed analysis suggests that the increase in anharmonic effects at higher temperatures/temperature
differences is mainly responsible for the diffusive transport regime in the longer chains.

\begin{acknowledgements}

We acknowledge financial support from the UK EPSRC, under Grant No. EP/J019259/1.

\end{acknowledgements}

\appendix

\section{Calculation of the heat current}
\label{app:Jheat}

For the systems considered here, the Hamiltonian of the central region is given by,
\begin{equation}
H_C = \sum_{i\alpha} \frac{1}{2m_i} p_{i\alpha}^2 + V\left(\{r_{i\alpha}\}\right) ,
\label{eq:H_C}
\end{equation}
where $p_{i\alpha}$ and $r_{i\alpha}$ are the momentum and position of the DOFs of
the central region.
The Hamiltonian $H_\nu$ of the harmonic bath $\nu=1,2$ is given by
\begin{equation}
H_\nu =\sum_{b_\nu}\frac{1}{2\mu_{l_\nu}} p_{b_\nu}^{2}  + V^{\rm harm}_{(\nu)}\left(\{u_{b_\nu}\}\right)
\label{eq:H_nu}
\end{equation}
with $p_{b_\nu}$ and $u_{b_\nu}$ being the the momentum and position of the DOFs of
the bath $\nu$, and $V^{\rm harm}_{(\nu)}$ is the harmonic potential energy of the bath.
The coupling between the bath $\nu=1,2$ and the central region is given by
\begin{equation}
H_{\nu C} = \sum_{b_\nu} \mu_{l_\nu }f_{b_\nu}(\{r_{i\alpha}\}) u_{b_\nu} ,
\label{eq:H_nuC}
\end{equation}
where the coupling force $f_{b_\nu}(\{r_{i\alpha}\})$ between the central system and the bath $\nu$
is arbitrary with respect to the central system DOF.

In Refs.~[\onlinecite{Kantorovich:2008,Stella:2014,Ness:2016}], it was shown that the equation of motion (EOM) 
of the DOFs in the bath are given by
\begin{equation}
\mu_{l_\nu} \ddot{u}_{b_\nu} = - \frac{\partial V^{\rm harm}_{(\nu)} } {\partial u_{b_\nu} }
-\mu_{l_\nu}f_{b_\nu}\left(\{r_{i\alpha}\}\right) .
\label{eq:bath_eom}
\end{equation}
The EOM for the DOF in the central region are given by
\begin{equation}
m_{i}\ddot{r}_{i\alpha} = 
-\frac{\partial V\left(\mathbf{r}\right)}{\partial r_{i\alpha}}
-\sum_{\nu=1}^2 \sum_{b_\nu} \mu_{l_\nu} g_{i\alpha,b_\nu} u_{b_\nu}
\label{eq:sys_eom}
\end{equation}
where $g_{i\alpha,b_\nu}=\partial f_{b_\nu}/\partial r_{i\alpha}$.

Now, we define the heat current $J_\nu$ flowing between the central region and the bath $\nu=1$
as follows:
\begin{equation}
J_\nu = \frac{d}{dt}\left( H_\nu + H_{\nu C} \right) .
\label{eq:Jheat_nu}
\end{equation}

This definition arises from the local continuity equation (between the Hamiltonian density and the
corresponding flux \cite{Lepri:2003}) integrated over the volume encompassing the bath $\nu$ DOFs.
The volume integration of the time derivative of the Hamiltonian density gives the RHS of
Eq.~(\ref{eq:Jheat_nu}). The volume integration of the divergence of the flux is transformed into
a surface integral of the flux over the interface between the bath $\nu$ and the central region.
The latter surface integral (with the surface normal vector pointing from the bath towards
the central region) provides the total heat current $J_\nu$ flowing through the interface between 
the bath $\nu$ and the central region.

Using elementary calculus ($dH(p,q)/dt = \dot{p}\ dH/dp + \dot{q}\ dH/dq)$
and from the definition of $H_\nu$ and $H_{\nu C}$, one finds that
\begin{equation}
\begin{split}
J_\nu = \sum_{b_\nu} \dot{u}_{b_\nu} 
\left( \mu_{l_\nu} \ddot{u}_{b_\nu}+ \frac{\partial V^{\rm harm}_{(\nu)} } {\partial u_{b_\nu} }
\right)
+ \mu_{l_\nu }f_{b_\nu} \dot{u}_{b_\nu} \\ \nonumber
+ \mu_{l_\nu }u_{b_\nu} \sum_{i\alpha} g_{i\alpha,b_\nu} \dot{r}_{i\alpha}
\end{split}
\label{eq:Jheat_nu_A}
\end{equation}

By using the EOM Eq.~(\ref{eq:bath_eom}) of the bath DOF, the above equation reduces to
\begin{equation}
\begin{split}
J_\nu = 
 \sum_{i\alpha} \dot{r}_{i\alpha}
\left( \sum_{b_\nu} \mu_{l_\nu }g_{i\alpha,b_\nu} u_{b_\nu}  \right) .
\end{split}
\label{eq:Jheat_nu_B}
\end{equation}

Note that, by definition, $\mu_{l_\nu }g_{i\alpha,b_\nu}$ is the spatial derivative of the
force between the DOFs $i\alpha$ and $b_\nu$, and $u_{b_\nu}$ are small displacements of the bath
DOFs around their equilibrium positions. Therefore the quantity $\mu_{l_\nu }g_{i\alpha,b_\nu}u_{b_\nu}$
is a force and its sum $\mathcal{F}_ {i\alpha}^{(\nu)}=\sum_{b_\nu} \mu_{l_\nu }g_{i\alpha,b_\nu}u_{b_\nu}$
can be seen as the total force acting on the DOF $i\alpha$ due to its coupling to the bath $\nu$.
Consequently the hear current can be expressed as the sum of the products of velocity times
force:
\begin{equation}
J_{\rm heat} = 
 \sum_{i\alpha} \dot{r}_{i\alpha} \mathcal{F}_ {i\alpha}^{(\nu)}
\label{eq:Jheat}
\end{equation}
For our 1D system, Eq.~(\ref{eq:Jheat}) simply reduces to 
$J_{\rm heat} = \sum_{i} \dot{x}_i  \mathcal{F}_ {ix}^{(\nu)}$.

In the steady-state, the current is supposed to be the same (up to
thermal fluctuations) in between any pair of atoms in the 1D chain.
In practice, we calculate the heat current between each pair $(n,n+1)$ 
of atoms contained in the 1D chains.
To make further connections with previous studies \cite{Lepri:2003,Segal:2003},
we use the following notations  $\dot{x}_n \equiv v_n$ and
$\mathcal{F}_ {nx}^{(\nu)} \equiv f_n$.
The heat current, between  the pair $(n,n+1)$, is given by 
$j_{n,n+1} = (v_n f_n + v_{n+1} f_{n+1} ) /2$,
and we average $j_{n,n+1}$ over all pairs
of atoms of the chain.

We perform a further average over time, in the
time range typically 200 to 300 ps, where the
system has reached the steady-state to get the steady state heat
current $J_{\rm heat}$.

Note that, when the force $f_n$ on atom $n$ in the chain is given by a 
(short-ranged) pairwise potential, it can be decomposed into two contributions
from the nearest-neighbours, i.e. $f_n=F_{n-1,n}+F_{n,n+1}$
with the nearest-neighbours forces $F_{n,m}=F_{m,n}$. 
Our definition of the heat current $\sum_n v_n f_n$ becomes then equivalent 
(after a few manipulation of the indices $n-1,n,n+1$) to the more commonly
used expression for the 
current $\sum_n (v_{n+1} + v_n )F_{n+1,n}$ 
usually found in the literature \cite{Lepri:2003,Segal:2003}.

\section{Heat current}
\label{app:Jheat_results}

\begin{figure}
\begin{centering}
\includegraphics[width=75mm]{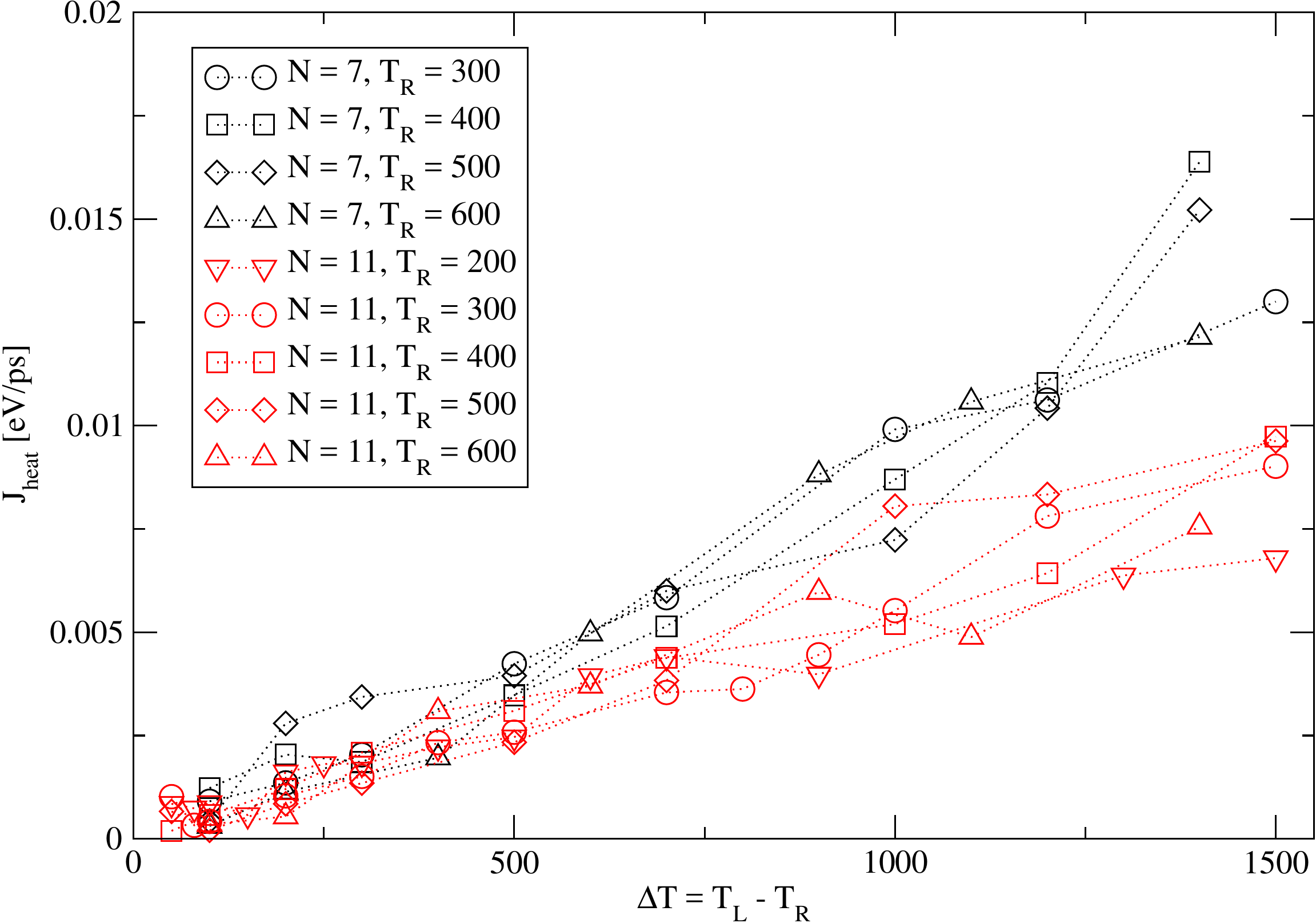}\\
\includegraphics[width=75mm]{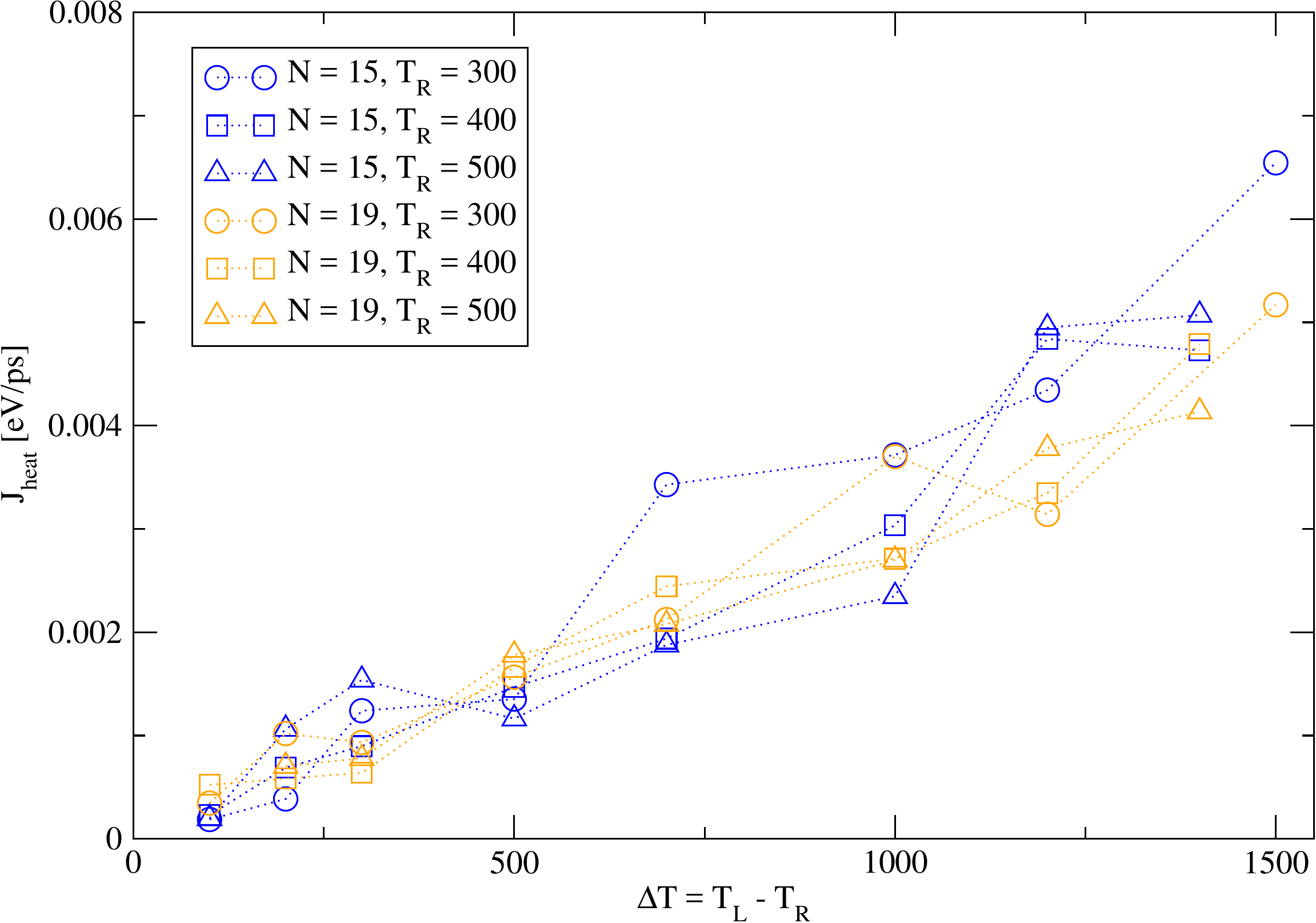}\\
\includegraphics[width=75mm]{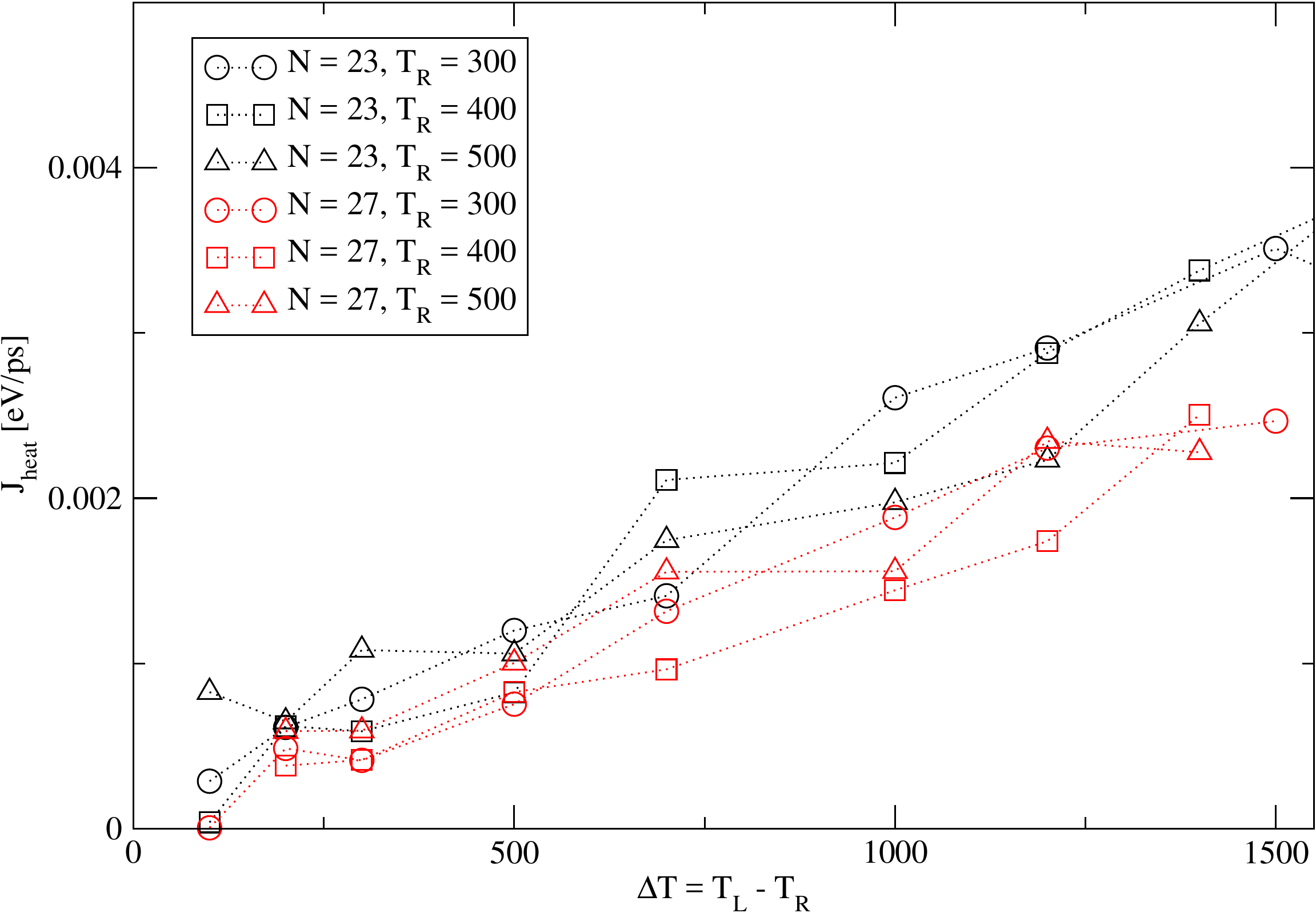}
\end{centering}
\caption{Compilation of \HN{most of the results} for the heat current $J_{\rm heat}$ through the
system, 1D chains of different length $N=7,11,15,23,27$ and for different 
baths' temperatures $T_L$ and $T_R$. Here the temperature difference is $\Delta T = T_L -T_R$.}
\label{fig:Jheat}
\end{figure}

For information, we present in Figure~\ref{fig:Jheat} most of the results we have obtained 
for the heat current $J_{\rm heat}$
flowing across all chains (length $N=7,11,15,19,23,27$) and for all sets of
temperatures $T_L$ and $T_R$.
From these results, it can be seen that the heat current increases with increasing
temperature differences $\Delta T = T_L -T_R$ (as expected). Furthermore, $J_{\rm heat}$
decreases when the length of the chain increases; such a behaviour is more pronounced
for large temperature differences $\Delta T$.

\HN{Furthermore, we can estimate an error on the numerical values calculated for 
the heat current $J_{\rm heat}$. 
We recall that the latter is calculated as an average of the heat current 
between  the pairs $(n,n+1)$ in the chain: 
\begin{equation}
J_{\rm heat}
=
\frac{1}{N_{\rm pair}} \sum_{n=1}^{N-1} \langle j_{n,n+1} \rangle_\tau 
= 
\frac{1}{N_{\rm pair}} \sum_{n=1}^{N-1} (
\langle v_n f_n \rangle_\tau + \langle v_{n+1} f_{n+1} \rangle_\tau ) / 2 \ .
\end{equation}
The local heat current $\langle v_n f_n \rangle_\tau$ is obtained from the time average
of $v_n(t) f_n(t)$ over the time period $t \in [t_{\rm start}, t_{\rm end}]$ with
$t_{\rm end}-t_{\rm start}=\tau$.

From the different simulations,
we estimate an absolute error of $\Delta(j_{n,n+1}) \sim 0.008$ [eV/ps] for the local
current. As the heat current $J_{\rm heat}$ is an average over the different pairs,
we assume that the corresponding standard error of the mean behaves as 
$\Delta(J_{\rm heat}) \sim 0.008 / \sqrt{N_{\rm pair}}$ [eV/ps].

Within this error margin, we can safely assume that the heat current calculated for
$T_L=400$ and $T_R=300$ and 
for 
$T_L=500$ and $T_R=400$
(shown in Fig.~\ref{fig:Jlin_vsL} by the black curves with circles and triangles respectivelly)
is nearly independent of the chain length.
}

\end{document}